\documentclass[a4paper,10pt]{article}
\pdfoutput=1
\usepackage[utf8]{inputenc}
\usepackage{amsmath}    
\usepackage{amssymb}             
\usepackage{todonotes}
\usepackage[ngerman,british]{babel}
\usepackage[left=20mm, right=20mm, top=25mm, bottom=25mm,bindingoffset=5mm, twoside]{geometry}
\usepackage{graphicx}
\usepackage{cite}
\usepackage{slashed}
\usepackage{xspace}

\newcommand{\xf}{X_t/M_S}

\newcommand{\DR}{{\overline{\text{DR}}}}
\newcommand{\MS}{{\overline{\text{MS}}}}
\newcommand{\OS}{{\text{OS}}}

\newcommand{\tgiu}{\tilde g_{1u}}
\newcommand{\tgid}{\tilde g_{1d}}
\newcommand{\tgiiu}{\tilde g_{2u}}
\newcommand{\tgiid}{\tilde g_{2d}}
\newcommand{\FHold}{\mbox{{\tt FeynHiggs2.11.3}}\xspace}
\newcommand{\FHnew}{\mbox{{\tt FeynHiggs2.12.0}}\xspace}
\newcommand{\FH}{\mbox{{\tt FeynHiggs}}\xspace}

\newcommand{\origttfamily}{}
\let\origttfamily=\ttfamily 
\renewcommand{\ttfamily}{\origttfamily \hyphenchar\font=`\-}

\begin{document}
 
\thispagestyle{empty}

\def\thefootnote{\fnsymbol{footnote}}

\begin{flushright}
MPP-2016-180
\end{flushright}

\vspace{2cm}

\begin{center}

{\Large {\bf Precise prediction for the light MSSM Higgs boson mass} }

\vspace{0.4cm}
 
{\Large {\bf combining effective field theory and fixed-order calculations}}

\vspace{1cm}

{\large {\sc
Henning Bahl}\footnote{email: hbahl@mpp.mpg.de} 
and
{\sc Wolfgang Hollik}\footnote{email: hollik@mpp.mpg.de}
}

\vspace*{.7cm}

{\large
{\sl
Max-Planck-Institut f\"ur Physik \\
(Werner-Heisenberg-Institut) \\[0.1cm]
F\"ohringer Ring 6, 
D--80805 M\"unchen, Germany
} }

\end{center}

\vspace*{2cm}

\begin{abstract}
In the Minimal Supersymmetric Standard Model heavy superparticles
introduce large logarithms in the calculation of the lightest
$\mathcal{CP}$-even Higgs boson mass. These logarithmic contributions
can be resummed using effective field theory techniques. For light
superparticles, however, fixed-order calculations are expected to be
more accurate. To gain a precise prediction also for intermediate mass
scales, both approaches have to be combined. Here, we report on an
improvement of this method in various steps: the inclusion of
electroweak contributions, of separate electroweakino and gluino
thresholds, as well as resummation at the NNLL level. These improvements can 
lead to significant numerical effects. In most cases, the lightest 
$\mathcal{CP}$-even Higgs boson mass is shifted downwards by about 1 GeV. 
This is mainly caused by higher order corrections to the $\MS$ top-quark mass. 
We also describe the implementation of the new contributions in the code {\tt FeynHiggs}.
\end{abstract}

\def\thefootnote{\arabic{footnote}}
\setcounter{page}{0}
\setcounter{footnote}{0}

\newpage

\section{Introduction}

With the discovery of the Higgs boson by the experiments ATLAS \cite{Aad:2012tfa} and CMS \cite{Chatrchyan:2012xdj} at the CERN Large Hadron Collider the Standard Model (SM) has been completed; there is, however, still ample room for Beyond Standard Model (BSM) physics. One of the best motivated and studied BSM models is the Minimal Supersymmetric Standard Model (MSSM) realizing the concept of supersymmetry (SUSY). It extends the Higgs sector of the SM by a second complex doublet leading to five physical Higgs particles ($h$, $H$, $A$ and $H^\pm$) and three (would-be) Goldstone bosons. The light $\mathcal{CP}$-even state $h$ can be identified with the discovered boson. At the tree-level, the Higgs sector can be conveniently parametrized by the mass of the $A$ boson, $M_A$, and the ratio of the vacuum expectation values of the two doublets, $\tan\beta = v_2/v_1$. 

So far, no direct hints for SUSY particles have been found. Still, the SUSY parameter space can be constrained indirectly by precision observables, with the Higgs-boson mass constituting an important precision observable on its own. Since the Higgs mass $M_h$ is very sensitive to quantum effects via loop contributions, much work has been dedicated to their calculation within the MSSM. The full one-loop result~\cite{Chankowski:1992er,Dabelstein:1994hb,Pierce:1996zz}, the dominant two-loop corrections~\cite{Heinemeyer:1998kz,Heinemeyer:1998jw,Heinemeyer:1998np,Heinemeyer:1999be,Heinemeyer:2004xw,Zhang:1998bm,Espinosa:1999zm,Degrassi:2001yf,Hempfling:1993qq,Brignole:2002bz,Dedes:2003km,Carena:1995wu,Casas:1994us,Brignole:2001jy,Espinosa:2000df} as well as partial three-loop results~\cite{Harlander:2008ju,Kant:2010tf} are known. For heavy SUSY particles,  fixed-order calculations suffer from large logarithms originating in a potentially huge hierarchy between the electroweak scale and the SUSY scale. Therefore, effective field theory (EFT) calculations have been developed to resum these logarithmic contributions~\cite{Haber:1993an,Carena:1995wu,Haber:1996fp}. Recent works have refined these methods to include gaugino/higgsino thresholds \cite{Binger:2004nn,Giudice:2011cg,Bagnaschi:2014rsa} and to allow for light non-standard Higgs particles~\cite{Lee:2015uza}. Furthermore, resummation at the next-to-next-to-leading logarithm (NNLL) level has been adressed in~\cite{Draper:2013oza,Bagnaschi:2014rsa,Vega:2015fna}.

These computations, however, do not capture the effect of terms that would be suppressed only in case of a heavy  SUSY scale. Thus, fixed-order calculations are expected to be more accurate for low SUSY scales. To gain the most accurate prediction for intermediate SUSY scales, both approaches have to be combined. This allows also to profit from the other advantages of the diagrammatic approach: the easy inclusion of many different SUSY scales, and the  full control over the Higgs boson self-energies, which are needed for other observables (e.g. production and decay rates).  The authors of~\cite{Hahn:2013ria} first realized the idea of combining the diagrammatic and the EFT approach and  implemented the method into the publicly available program \FH \cite{Heinemeyer:1998yj,Heinemeyer:1998np,Hahn:2009zz,Degrassi:2002fi,Frank:2006yh,Hahn:2013ria}, which also contains the complete fixed-order one-loop result as well as dominant two-loop results; NLL resummation was done for the strong and top Yukawa coupling 
enhanced logarithmic terms beyond the two-loop order.
Here, we report on an extension of this work in a threefold respect: the inclusion of the electroweak contributions, 
the inclusion of separate electroweakino and gluino thresholds, and resummation of logarithms proportional to the top Yukawa coupling and the strong gauge coupling at the NNLL level.

In Section~\ref{SecEFTCalculation}, we outline  the EFT calculation, focusing on the ingredients needed to include electroweak contributions, gaugino/higgsino thresholds and NNLL resummation. In Section~\ref{SecCombining}, we describe how the result of the EFT calculation is consistently combined with the fixed-order diagrammatic result. In Section~\ref{SecFeynHiggs}, we discuss the implementation of the improvements in {\tt FeynHiggs}. In Section~\ref{SecNumericalAnalysis}, we present a numerical analysis showing the impact of the improved version on the calculation of $M_h$, with conclusions in Section~\ref{SecConclusions}.

\section{Effective field theory calculation}\label{SecEFTCalculation}

In the case of heavy SUSY particles, large logarithms appear in explicit diagrammatic calculations 
making a fixed-order calculation an unreliable tool.
The origin of this problem is the large hierarchy between the electroweak scale and the SUSY scale. Effective field theory techniques allow to resum these large logarithms to all orders and thus get stable predictions.

In the simplest EFT framework, all SUSY particles are assumed to share a common mass scale $M_S$ (this assumption will be relaxed below), where $M_S$ is the stop mass scale,
\begin{align}
M_S = \sqrt{m_{\tilde t_1}m_{\tilde t_2}},
\end{align} 
and $m_{\tilde t_{1,2}}$ are the stop masses. This scale is furthermore assumed to be much heavier than the electroweak scale. Below $M_S$, all SUSY particles are integrated out from the full theory. Thus, the low energy effective theory below the SUSY scale is the SM. 

The effective couplings of the EFT
are fixed by matching to the full MSSM at the matching scale $M_S$
(in the simplest case of an effective SM below $M_S$, 
this concerns only the Higgs self-coupling $\lambda$). 
All of the other couplings of the EFT are fixed by matching them to physical observables at the electroweak scale \cite{Buttazzo:2013uya}, 
e.g.~the top Yukawa coupling is extracted from the top-quark pole mass.

In the EFT framework, we calculate the Higgs mass via the relation
\begin{align}\label{EFTresult}
\left(M^\text{EFT}_{h}\right)^2 = \lambda(M_t)v^2,
\end{align} 
with the self-coupling 
$\lambda$ evaluated at the top-quark pole mass $M_t$ and with the electroweak vacuum expectation value $v$ (see \cite{Carena:2000dp} and references therein). Since all SUSY particles are integrated out at the electroweak scale, this ensures that all large logarithms which would appear explicitly in a diagrammatic calculation in the full model framework are contained in $\lambda(M_t)$. $\lambda(M_t)$ is obtained 
via renormalization group equations (RGEs) from $\lambda(M_S)$,  which is determined by matching $\lambda$ to the full MSSM at $M_S$.
The running between $M_S$ and $M_t$ corresponds to a resummation of large logarithms.

For the resummation of leading logarithms (LL), one-loop RGEs and tree-level matching conditions are needed; for the resummation of leading and next-to-leading logarithms (NLL), two-loop RGEs and one-loop matching conditions, and, accordingly,
for the resummation of leading, next-to-leading and next-to-next-to-leading logarithms, three-loop RGEs and two-loop matching conditions.

\subsection{Electroweak contributions}

As a first improvement with respect to \cite{Hahn:2013ria}, we include electroweak contributions in the resummation procedure at the NLL level. Correspondingly, we use the full two-loop RGEs of the SM (see \cite{Buttazzo:2013uya} and references therein), including terms proportional to the electroweak gauge couplings $g$ and $g'$ (for $SU(2)$ and $U(1)$), to evolve the SM couplings. 

Furthermore, the threshold correction of the Higgs self-coupling at the SUSY scale has to be extended at the one-loop level by adding  
the various electroweak one-loop contributions,
\begin{align}\label{threshold1L}
\lambda_{\text{SM}}(M_S) = \frac{1}{4}(g^2+g'^2) \cos^2(2\beta)+\Delta_{\text{stop}}\lambda+\Delta_{\text{heavyH}}\lambda+\Delta_{\text{EWino}}\lambda+\Delta_{\DR\rightarrow\MS}\lambda.
\end{align}
$\Delta_{\text{stop}}\lambda$ is the contribution from the top and stop sector (extended by electroweak contributions in comparison to \cite{Hahn:2013ria}); $\Delta_{\text{heavyH}}\lambda$, the contribution from the heavy non-SM Higgs bosons; $\Delta_{\text{EWino}}\lambda$ the contribution from charginos and neutralinos. The term $\Delta_{\DR\rightarrow\MS}\lambda$ accounts for the fact that the tree-level contribution is expressed in terms of $\MS$-renormalized gauge couplings of the SM and not in terms of $\DR$-renormalized gauge couplings of the MSSM. All of these threshold corrections have been derived in previous works \cite{Haber:1993an,Giudice:2011cg,Carena:1995wu,Bagnaschi:2014rsa}. We use the expressions given in \cite{Bagnaschi:2014rsa}. Accordingly, also the relations used to extract SM gauge and Yukawa couplings from physical observables at $M_t$ must include electroweak one-loop corrections \cite{Buttazzo:2013uya}. This is especially relevant for the $\MS$ top-quark mass, respectively the top Yukawa coupling, as will be discussed later in the section on results.

\subsection{Gaugino--higgsino thresholds}

The assumption of a common mass scale for all SUSY particle is quite limiting. To allow for electroweakinos (charginos and neutralinos) lighter than $M_S$ (but still much heavier than the electroweak scale), we introduce an additional electroweakino threshold $M_\chi$. We assume that all charginos and neutralinos are nearly mass degenerate (having the mass $M_\chi$), 
\begin{align}
M_\chi = M_1 = M_2 = \mu\text{ with }M_Z\ll M_\chi \le M_S,
\end{align}
where $M_1$ and $M_2$ are the electroweak gaugino soft-breaking masses and $\mu$ is the Higgsino mass parameter.

This means that at $M_S$ all SUSY particles but charginos and neutralinos are integrated out. The corresponding EFT below $M_S$, the split model, is the SM with charginos and neutralinos added. The corresponding effective Lagrangian reads \cite{Bagnaschi:2014rsa}
\begin{align}
\mathcal{L}_{\text{split}} =&\mathcal{L}_{\text{SM}} + ... -\frac{1}{2}M_\chi \tilde W \tilde W -\frac{1}{2}M_\chi \tilde B \tilde B - M_\chi \tilde{\mathcal{H}}_2\cdot\tilde{\mathcal{H}}_1 \nonumber\\
&- \frac{1}{\sqrt{2}}H^\dagger\left(\tgiiu\sigma^a\tilde W^a+\tgiu\tilde B\right)\tilde{\mathcal{H}}_u - \frac{1}{\sqrt{2}}H\cdot\left(-\tgiid\sigma^a\tilde W^a+\tgid\tilde B\right)\tilde{\mathcal{H}}_d + h.c. 
\end{align}
The bino field is denoted by $\tilde B$, the wino field by $\tilde W$ and the higgsino fields by $\tilde{\mathcal{H}}_{u,d}$. The ellipsis stands for the associated kinetic terms. The effective Higgs-Higgsino-gaugino couplings are labeled $\tilde g_{1u,...}$. The number in the subscript refers to the symmetry group $U(1)$ or $SU(2)$, the letter to the involved Higgs doublet. These effective couplings are determined by a one-loop matching of the split model to the full MSSM at the scale $M_S$ (for explicit expressions, see \cite{Giudice:2011cg,Bagnaschi:2014rsa}). All couplings are evolved between the electroweakino scale and the stop mass scale using two-loop split model RGEs, which can be found in \cite{Binger:2004nn,Giudice:2011cg,Bagnaschi:2014rsa}.

At the scale $M_\chi$, all electroweakinos are integrated out, and the remaining EFT below $M_\chi$ is the SM. We match the SM to the split model using the threshold corrections given in \cite{Giudice:2011cg,Bagnaschi:2014rsa}, i.e. the term $\Delta_{\text{EWino}}\lambda$ in Eq.~(\ref{threshold1L}) is now part of the matching condition of $\lambda$ at $M_\chi$. Also the top Yukawa coupling receives a threshold correction at the electroweakino scale. Below $M_\chi$ the SM RGEs are used for evolving the couplings.

In addition to allowing for light charginos and neutralinos, we also consider the case of a light gluino. This case is implemented by introducing an additional threshold marked by the gluino mass $M_{\tilde g}$, below which the gluino is integrated out. The gluino is also assumed to be much heavier than the electroweak scale such that eventually the SM is recovered as the EFT close to the electroweak scale. However, no assumption about the ordering of $M_{\tilde g}$ and $M_\chi$ is made, i.e. $M_{\tilde g}\le M_\chi$ as well as $M_{\tilde g}> M_\chi$ is allowed. Since the gluino does not couple directly to the Higgs boson, no additional one-loop matching condition for $\lambda$ has to be considered. The same argument applies for the electroweak gauge couplings, the Yukawa couplings (in the absence of sfermions) and the effective Higgs-Higgsino-gaugino couplings of the split model. An explicit calculation also shows that the strong gauge coupling does not receive a threshold correction. However, the presence of the gluino in the EFT above $M_{\tilde g}$ modifies the RGEs (see App.~\ref{GluinoRGE}).

\subsection{NNLL resummation}

As a further improvement, we include resummation at the NNLL level. This is restricted to the dominating contributions resulting from 
the top Yukawa coupling  $y_t$, respectively $\alpha_t=y_t^2/4\pi$, and  the strong gauge coupling $g_3$, respectively  $\alpha_s=g_3^2/4\pi$.
NNLL resummation requires two-loop threshold corrections. 
Therefore, we extend Eq.~(\ref{threshold1L}) by the corresponding two-loop contributions,
\begin{align}
\lambda_{\text{SM}}(M_S) =\frac{1}{4}(g^2+g'^2) \cos^2(2\beta)+\Delta_{\text{stop}}\lambda+\Delta_{\text{heavyH}}\lambda+\Delta_{\text{EWino}}\lambda+\Delta_{\DR\rightarrow\MS}\lambda+\Delta_{\alpha_s\alpha_t}\lambda+\Delta_{\alpha_t^2}\lambda.
\end{align}
These terms have already been calculated based on the work of \cite{Espinosa:2000df}. The $\mathcal{O}(\alpha_s\alpha_t)$ corrections are given in \cite{Draper:2013oza} and \cite{Bagnaschi:2014rsa}; the pure top Yukawa correction $\mathcal{O}(\alpha_t^2)$ are listed in \cite{Draper:2013oza} and in a slightly different form in \cite{Vega:2015fna}. We take use of the expressions given in \cite{Vega:2015fna}. 

Also the matching conditions for the SM gauge and Yukawa couplings at $M_t$ have to be extended to include the $\mathcal{O}(\alpha_s^2,\alpha_s\alpha_t,\alpha_t^2)$ corrections. These are taken from~\cite{Buttazzo:2013uya}. 
The matching condition for the top Yukawa coupling involves the $\MS$ top-quark mass
which for NNLL resummation is obtained from the pole mass by means of the standard 
QCD  and top Yukawa corrections at the two-loop level~\cite{Buttazzo:2013uya}. 
%

Furthermore, three-loop RGEs are needed for the coupling constant evolution. Since only NNL logarithms of $\mathcal{O}(\alpha_s,\alpha_t)$ are resummed in this step, we neglect the electroweak gauge couplings at the three-loop level of the needed RGEs. All couplings of electroweakinos, being present below $M_S$ for $M_\chi<M_S$, are proportional to the electroweak gauge couplings when their matching conditions at $M_S$ are plugged in. In consequence, their presence has no influence on the form of the three-loop RGEs at this level of approximation. Hence for all considered hierarchies at all scales below $M_S$, the needed three-loop RGEs are just the corresponding SM RGEs, which are well known \cite{Mihaila:2012fm,Mihaila:2012pz,Chetyrkin:2012rz,Bednyakov:2012en,Bednyakov:2012rb,Bednyakov:2013eba,Chetyrkin:2013wya}. The same argument implies that the two-loop matching conditions of $\lambda$ do not have to be modified for $M_\chi$ lower than $M_S$.

For NNLL resummation, we have to restrict ourselves to the case of $M_{\tilde g}$ equal to $M_S$ in the resummation procedure, since three-loop RGEs for the SM with added gluino are not known. Nevertheless, the numerical effect of a gluino threshold is so small that it can be safely neglected, as will be seen  in the numerical results.

\section{Combining fixed-order and EFT calculations}
\label{SecCombining}

The final prediction for the physical Higgs mass is obtained by 
adding the fixed-order result achieved by a Feynman-diagrammatic calculation
and the result obtained from the effective field theory method 
with appropriate subtractions in order to avoid double-counting of terms
already contained in the fixed-order expressions
(see also \cite{Hahn:2013ria}),
\begin{align}\label{MasterEquation}
M_h^2 =& \left(M_h^2\right)^\text{FD}(X_t^\OS) + \Delta M_h^2,\\
\Delta M_h^2 =& \left(M_h^2\right)^\text{EFT}(X_t^\DR) - \left(\Delta M_h^2\right)^{\text{1L,2L logs}}(X_t^\OS) \nonumber\\
& - \left(\Delta M_h^2\right)^{\text{EFT, non-log}}(X_t^\OS),
\end{align}
where $X_t$ is the mixing parameter in the stop squared-mass matrix. $\left(M_h^2\right)^\text{FD}$ denotes the one- and two-loop fixed-order Feynman-diagrammatic result in the on-shell renormalization scheme, 
as implemented in \FH. 
$\Delta M_h^2$ is the result of the resummation beyond two-loop order, which consists of the result of the EFT calculation $\left(M_h^2\right)^\text{EFT}$ together with the proper subtraction terms.
$\left(M_h^2\right)^\text{EFT}$ is obtained via Eq.~(\ref{EFTresult}) from the RGEs and threshold corrections
involving the SUSY parameter $X_t$ defined in the $\DR$-scheme at the scale~$M_S$. 

The first subtraction term
$\left(\Delta M_h^2\right)^{\text{1L,2L logs}}$ ensures that the one- and two-loop logarithms in the OS scheme, already contained in the Feynman-diagrammatic result, are not counted twice. We extracted these logarithms in the EFT framework by solving the system of RGEs iteratively and converting to the OS scheme afterwards. As a cross-check, we also identified the one-loop logarithms within the Feynman-diagrammatic result finding agreement (see \ref{AppExplicitExpr} for explicit expressions). It should be noted that {\tt FeynHiggs} also allows to choose a $\MS$ top-quark mass \cite{Heinemeyer:1998np}. If this option is switched on, we have to subtract the one- and two-loop logarithms as contained in the Feynman-diagrammatic result, i.e.\ as obtained with a $\MS$ top-quark mass. 

The second subtraction term $\left(\Delta M_h^2\right)^{\text{EFT, non-log}}$ is introduced to cancel all non-logarithmic terms contained in the EFT result. They originate from the matching conditions of the Higgs self-coupling and have to be subtracted when only higher-order logarithmic contributions are added to the Feynman-diagrammatic result. 

A particular  issue to be taken care of when combining the diagrammatic result with the EFT calculation, is the choice of the renormalization scheme. The EFT calculation uses minimal subtraction schemes ($\DR$ for scales above $M_S$, $\MS$ for scales below $M_S$) for renormalization. In contrast, in the diagrammatic calculation a mixed OS/$\DR$ scheme is employed (see \cite{Frank:2006yh} for a detailed description). Consequently, the input parameters of the EFT calculation are $\MS$/$\DR$ parameters, whereas they are OS parameters in the diagrammatic calculation \cite{Carena:2000dp}, as indicated in Eq.~(\ref{MasterEquation}). The logarithmic subtraction term takes OS parameters as input, because we want to avoid double-counting of the one- and two-loop logarithms in the OS scheme. Also the non-logarithmic subtraction term takes OS parameters as input, although the non-logarithmic terms contained in the EFT result are parametrized with $\DR$ parameters. 
This is owing to the fact that non-logarithmic terms in the $\DR$ scheme lead to logarithmic terms in the OS scheme;
consequently, the OS two-loop logarithms of the Feynman-diagrammatic result would not be reproduced
when $\DR$ parameters were used as input.

We choose to work with OS parameters as principal input. This means that OS input parameters are converted to $\DR$ parameters when used as input for the EFT calculation. We restrict ourselves to a one-loop conversion involving only terms proportional to large logarithms. This conversion is sufficient to reproduce all large logarithms already contained in the diagrammatic two-loop result of \FH. In contrast, non-logarithmic terms and higher loop-order terms would lead to terms in the EFT result which correspond to unknown higher-order corrections in an OS renormalized diagrammatic result. We however intend to add the resummed logarithms as obtained in the $\MS$/$\DR$ scheme to the diagrammatic result. In consequence, all terms beyond one-loop logarithms have to be omitted.

The main input parameters of the EFT calculation are the stop mass scale $M_S$ and the stop mixing parameter $X_t$. The conversion of $M_S$ does not involve any large logarithms \cite{Espinosa:2000df,Hahn:2013ria}; hence, $M_S$ is not converted. The conversion of $X_t$ at $\mathcal{O}(\alpha_s,\alpha_t)$ is given by\footnote{The $X_t^2$ term is missing in~\cite{Hahn:2013ria}, but it is properly included in \FH and in agreement with~\cite{Espinosa:2000df}.} 
\begin{align}\label{Xt_conversion_formula}
X_t^\DR = X_t^\OS \left[1 + \left(\frac{\alpha_s}{\pi} -\frac{3 \alpha_t}{16\pi}(1-X_t^2/M_S^2)\right)\ln \frac{M_S^2}{M_t^2}\right].
\end{align}

The only other input parameters in our EFT calculation are the intermediate electroweakino mass scale~$M_\chi$ and the gluino mass $M_{\tilde g}$. Since the diagrammatic \FH result so far contains two-loop corrections only in the gaugeless limit, a conversion of $M_\chi$, which would contain only terms proportional to the electroweak gauge couplings, is not needed. Since the gluino mass appears first at the two-loop level, also a conversion of $M_{\tilde g}$ is not necessary.

A further issue to be discussed is the treatment of $\tan\beta$. In the EFT approach, $\tan\beta$ appears only in the matching condition of $\lambda$ at the SUSY scale $M_S$. This means that the $\DR$-renormalized $\tan\beta(M_S)$ is required as an input of the EFT calculation. In the Feynman-diagrammatic calculation, $\tan\beta$ is also a $\DR$-renormalized quantity. In \FH, the corresponding renormalization scale, however, is chosen to be $M_t$ and not $M_S$ \cite{Frank:2006yh}. In consequence, we need to relate $\tan\beta(M_t)$, which is used as input of the overall calculation, to $\tan\beta(M_S)$. This presents a problem, since there is no proper way to define $\tan\beta$ in the EFT below $M_S$, where the non SM-like Higgs bosons are integrated out. This problem has already been noted in \cite{Haber:1993an}. We find that without a running of $\tan\beta$ the EFT calculation does not reproduce the one-loop result of the Feynmann diagrammatic calculation (see \ref{AppExplicitExpr} for explicit expressions). This strongly motivates to evolve $\tan\beta$ between $M_t$ and $M_S$ despite the lack of a rigoros definition. In practice, we regard $\tan\beta$ as a high-energy parameter with an evolution according to the one-loop RGE of the  
MSSM~\cite{Haber:1993an},
\begin{align}
\frac{1}{\tan^2\!\beta}\, \frac{d\tan^2\!\beta}{d\ln Q^2}=-\frac{3}{16\pi^2}h_t^2 ,
\end{align}
which is determined by the anomalous dimensions of the Higgs fields, with contributions only from the top-quark loops.
The parameter $h_t$ denotes the MSSM top Yukawa coupling, which at lowest order is related to the SM top Yukawa coupling $y_t$ by
\begin{align}
y_t = h_t \sin\beta.
\end{align}
Rewriting the RGE in terms of $y_t$ yields
\begin{align}
\frac{1}{1+\tan^2\!\beta}\, \frac{d\tan^2\!\beta}{d\ln Q^2}=-\frac{3}{16\pi^2} y_t^2.
\end{align}
Since only SM entries contribute to the running~\cite{Haber:1993an}, the RGE has not to be modified for scales below $M_S$, even if passing an intermediate  threshold. 
This method reproduces correctly the one-loop result of the diagrammatic calculation, 
as given in \ref{AppExplicitExpr}.
In principle, for a NLL resummation also the two-loop RGE should be employed, which for the MSSM can be found in \cite{Sperling:2013eva,Sperling:2013xqa}. It is, however, unclear which contributions of the two-loop RGE are due to SM particles and which are due to their supersymmetric partners. From a practical point of view,
numerical checks suggest that the two-loop running is negligible. Therefore, only the one-loop RGE is used in this work.

\section{Implementation in \FH}\label{SecFeynHiggs}

As explained in \cite{Hahn:2013ria}, the shift $\Delta M_h^2$ is implemented in \FH by adding it with a factor $1/\sin^2\beta$ to the $\phi_2\phi_2$ self-energy ($\phi_2$ is the $\mathcal{CP}$-even neutral component of the second Higgs doublet). In this way, the result of the resummation procedure enters also the calculation of other observables that are available from \FH.

The improved resummation of large logarithms is available in \FH from version {\tt 2.12.0} on. In this version, the new flag {\tt loglevel} is introduced to control the resummation procedure. The various options are
\begin{itemize}
\item
{\tt loglevel=0}: no resummation;
\item
{\tt loglevel=1}: $\mathcal{O}(\alpha_s,\alpha_t)$ LL and NLL resummation \\ (corresponds to former {\tt looplevel=3});
\item
{\tt loglevel=2}: full LL and NLL resummation; 
\item
{\tt loglevel=3}: full LL, NLL and $\mathcal{O}(\alpha_s,\alpha_t)$ NNLL resummation.
\end{itemize}
For {\tt loglevel} greater than one, electroweak NLO corrections to the $\MS$ top-quark mass are switched on automatically. 

So far, all matching conditions are only implemented for degenerate soft-breaking masses, meaning that all soft-breaking masses are set equal to their corresponding threshold scale. The diagrammatic part of the calculation, however, captures the effects of non-degeneracy in an exact way at the one- and two-loop level. The matching condition will be extended to the non-degenerate case in a future update to \FH.

\section{Numerical analysis}\label{SecNumericalAnalysis}

\begin{figure}\centering
\begin{minipage}{.495\textwidth}\centering
\includegraphics[scale=.93]{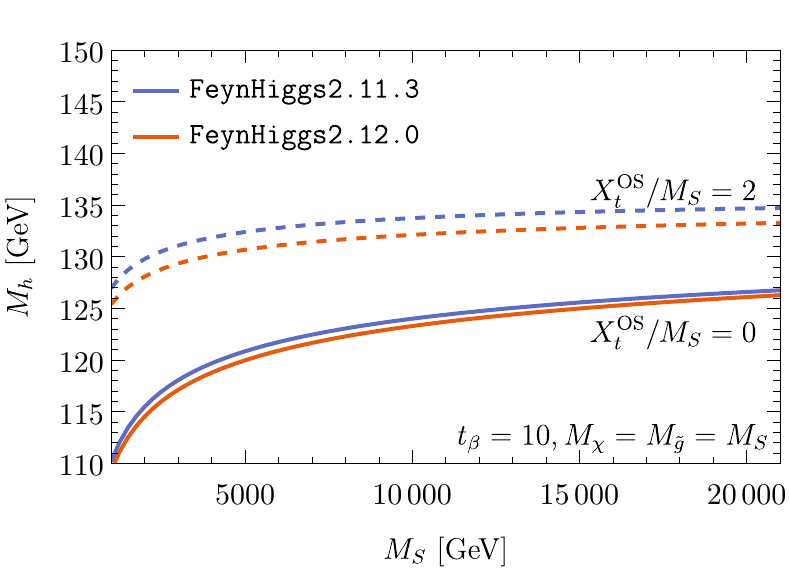}
\end{minipage}
\hfill
\begin{minipage}{.495\textwidth}\centering
\includegraphics[scale=.93]{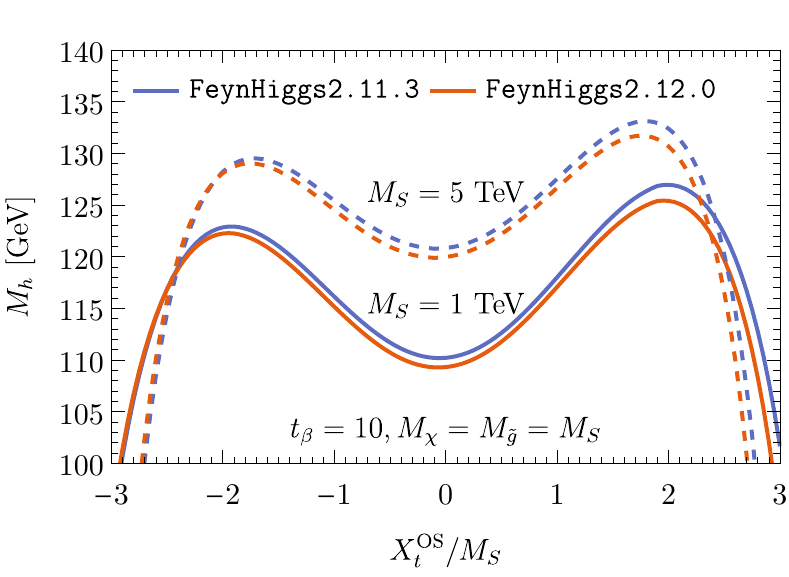}
\end{minipage}
\caption{Left: $M_h$ as a function of $M_S$ for $\xf=0$ (solid) and $\xf=2$ (dashed). The results of \FHold (blue) are compared to the results of \FHnew (red). Right: $M_h$ as a function of $X_t/M_S$ for $M_S=1$ TeV (solid) and $M_S=5$ TeV (dashed). The results of \FHold (blue) are compared to the results of \FHnew (red).}
\label{FigFHcomp}
\end{figure}

To analyse the numerical impact of the improved resummations, we first compare the results of the previous \FH version {\tt 2.11.3} with the new version {\tt 2.12.0}. As an example case, we look at a scenario where all soft-breaking masses as well as the \mbox{Higgsino} mass parameter are chosen to be equal to $M_S$, together with $t_\beta \equiv \tan\beta=10$. The results of \FHold are obtained with switched on $\mathcal{O}(\alpha_s,\alpha_t)$ LL and NLL resummation. Also the two-loop QCD correction to the $\MS$ top-quark mass are enabled, although no NNLL resummation is performed.
This is done because of the large numerical impact of this two-loop  correction on the Higgs mass calculation
(see the discussion at the end of this section).
For the results of \FHnew, all improvements discussed above are activated ({\tt loglevel=3}).

The comparison is displayed in Fig.~\ref{FigFHcomp}, where
the left panel shows $M_h$ as a function of $M_S$ for unmixed squarks with $\xf=0$ and for the mixed case with $\xf=2$.  For vanishing stop mixing, we observe a small downwards shift of $\lesssim 0.8$ GeV over the whole $M_S$ range, and a bit more for $X_t/M_S=2$, of $\lesssim 1.7$ GeV. The right panel in Fig.~\ref{FigFHcomp} shows $M_h$ as a function of $X_t/M_S$ for $M_S =1$ TeV and $M_S=5$ TeV. We observe a smaller shift for negative values of $X_t$; e.g.\ for $M_S=1$ TeV the difference between \FHold and \FHnew is $\sim 0.6$ GeV for $X_t/M_S=-2$, whereas it amounts to $\sim 1.6$ GeV for $X_t/M_S=2$.

\begin{figure}\centering
\begin{minipage}{.495\textwidth}\centering
\includegraphics[scale=.93]{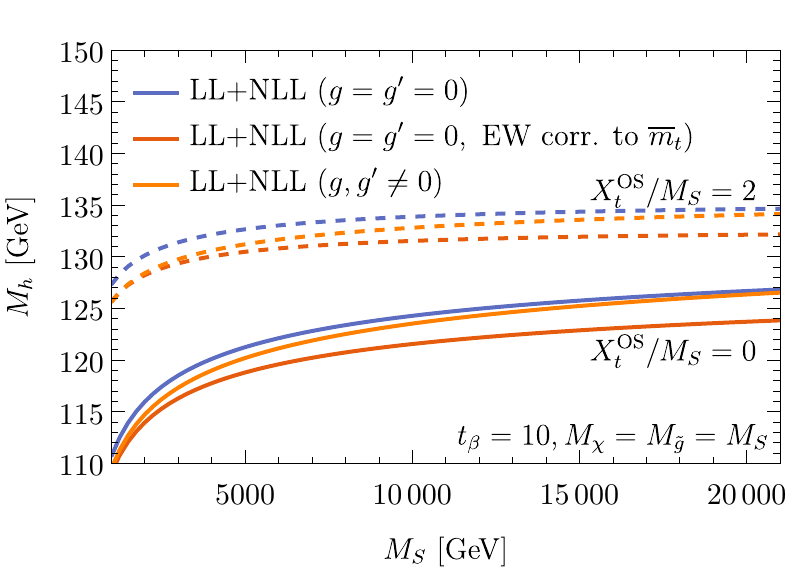}
\end{minipage}
\hfill
\begin{minipage}{.495\textwidth}\centering
\includegraphics[scale=.93]{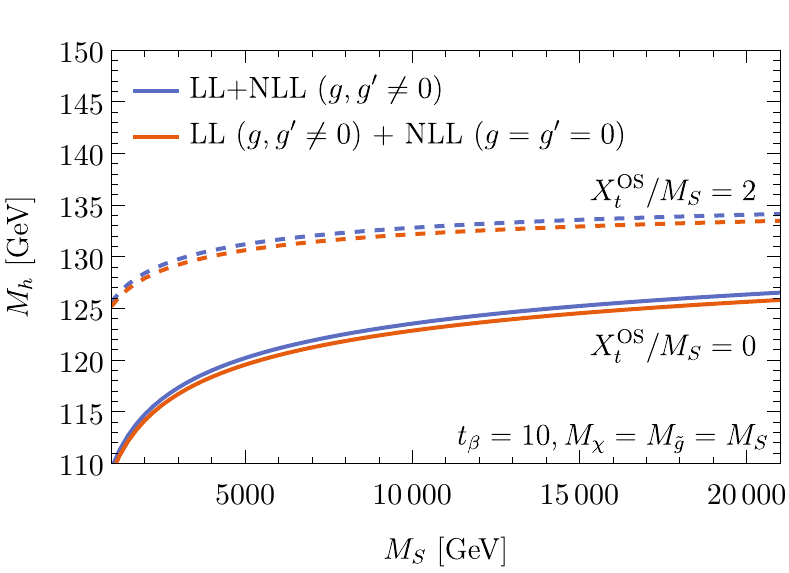}
\end{minipage}
\caption{Left: $M_h$ as a function of $M_S$ for $\xf=0$ (solid) and $\xf=2$ (dashed). The results with (orange) and without (blue) resummation of electroweak logarithms (LL+NLL) are compared. Furthermore, the result without resummation of electroweak logarithms but with electroweak NLO corrections to the $\MS$ top-quark mass (red) are shown. Right: The results with resummation of electroweak logarithms at the LL and NLL level (blue) and at the LL level only (red) are compared.}
\label{FigEWContr}
\end{figure}

To explore the origin of these shifts, we examine first the contribution of the resummation of logarithms proportional to the electroweak gauge couplings. The left panel of Fig.~\ref{FigEWContr} shows $M_h$ as a function of $M_S$ for $\xf=0$ and $\xf=2$. The results with a resummation of logarithms proportional to the electroweak gauge couplings ({\tt loglevel=2}) and without such a resummation are compared ({\tt loglevel=1}). The latter corresponds, apart from some minor fixes, to the result of \FHold. Furthermore, the result without resummation of logarithms proportional to the electroweak gauge couplings but with electroweak NLO corrections to the $\MS$ top mass is shown. For vanishing stop mixing, we observe a downwards shift of $\sim 1.2$ GeV for $M_S = 1$ TeV. This shift is almost completely caused by the electroweak NLO corrections to the $\MS$ top mass yielding a reduction of the $\MS$ top mass by 1.1 GeV. This translates directly to a downwards shift of $M_h$ \cite{Heinemeyer:1999zf}. For rising $M_S$, the downwards shift caused by the corrections to the $\MS$ top mass is more and more compensated by the upwards shift caused by the resummed logarithms proportional to the electroweak gauge couplings. For $X_t/M_S=2$, the behavior is very similar. For $M_S=1$ TeV, the downwards shift is larger ($\sim 1.7$ GeV) owing to the increased dependence on the $\MS$ top mass for nearly maximal stop mixing. For rising $M_S$, this downwards shift is again more and more compensated by the positive contributions of the resummed electroweak logarithms.  

The right panel of Fig.~\ref{FigEWContr} shows $M_h$ as a function of $M_S$ for $\xf=0$ and $\xf=2$. The results with a resummation of logarithms proportional to the electroweak gauge couplings at the LL and NLL level ({\tt loglevel=2}) and with a resummation of logarithms proportional to the electroweak gauge couplings at the LL level and vanishing electroweak gauge couplings at the NLL level are compared. We observe that the effect of a NLL resummation of electroweak logarithms is $\lesssim 0.5$ GeV over the whole $M_S$ range for both vanishing and nearly maximal mixing. This shows the minor importance of the electroweak NLL resummation in comparison to electroweak LL resummation, which leads to shifts of up to $2.5$ GeV for $M_S\sim 20$ TeV. 

\begin{figure}\centering
\begin{minipage}{.495\textwidth}\centering
\includegraphics[scale=.93]{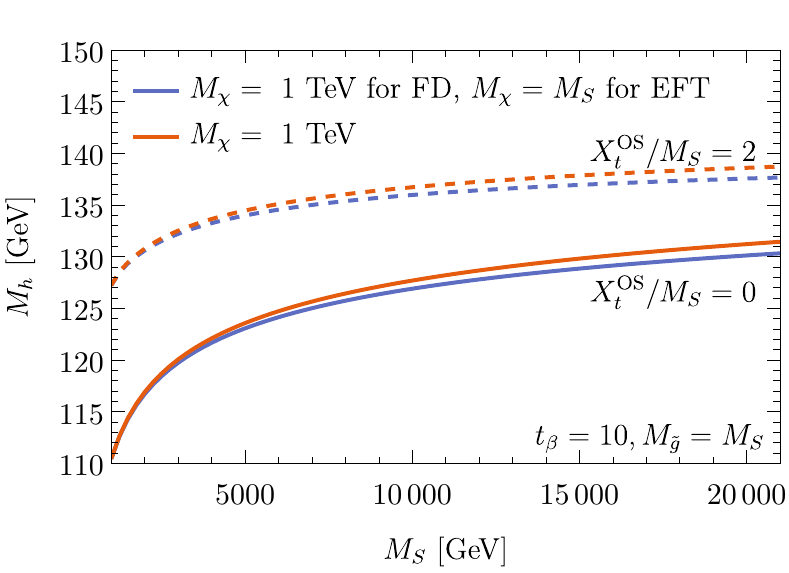}
\end{minipage}
\hfill
\begin{minipage}{.495\textwidth}\centering
\includegraphics[scale=.93]{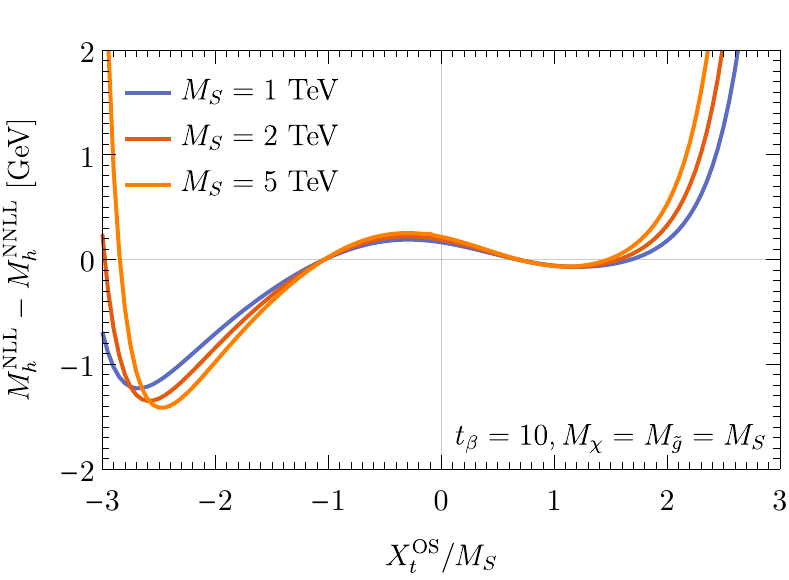}
\end{minipage}
\caption{Left: $M_h$ as a function of $M_S$ for $\xf=0$ (solid) and $\xf=2$ (dashed). The results with (red) and without (blue) electroweakino threshold are compared. Right: The difference between the NLL and the NNLL result as a function of $\xf$ for $M_S=1$ TeV (blue), $M_S=2$ TeV (red) and $M_S=5$ TeV (orange) is shown.}
\label{FigEWinoNNLLContr}
\end{figure}

The effect of the electroweakino threshold is investigated in the left panel of Fig.~\ref{FigEWinoNNLLContr}, which displays $M_h$ as function of $M_S$ for $\xf=0$ and $\xf=2$. In contrast to the previous figures, the electroweakino mass scale $M_\chi$ is not chosen to be equal to $M_S$, but is fixed to 1 TeV. To disentangle the effect of the electroweakino threshold in the EFT calculation from the fixed-order one-loop corrections due to neutralinos and charginos, we compare the results with a electroweakino threshold to the results without a separate electroweakino threshold. To get the results without a separate electroweakino threshold, we set $M_\chi=M_S$ in the EFT calculation (namely in $\Delta M_h^2$), but keep $M_\chi=1$ TeV in the Feynman-diagrammatic calculation. The plot clearly shows that the implementation of a separate electroweakino threshold becomes only relevant for $M_S\gtrsim 5$ TeV. This behavior does not depend on the size of the stop mixing.

The effect of a separate gluino threshold is found to be negligible. For $M_S$ between 1 TeV and 20 TeV, its inclusion shifts $M_h$ downwards by at most 0.2 GeV for $|X_t/M_S|\le 2$. The diagrammatic two-loop corrections capture almost the entire effect of varying $M_{\tilde g}$, which can be sizeable ($\sim 2$ GeV) for maximal mixing. This justifies to set $M_{\tilde g}=M_S$ in the resummation procedure in the case of NNLL resummation, as explained in Section~\ref{SecEFTCalculation}.

In the right panel of Fig.~\ref{FigEWinoNNLLContr}, the difference between the results without ({\tt loglevel=2}) and with ({\tt loglevel=3}) NNLL resummation as a function of $X_t/M_S$ is shown for $M_S=1$ TeV, $M_S=2$ TeV and $M_S=5$ TeV. Between $X_t/M_S \sim -1$ and $X_t/M_S \sim 1.5$, we observe only small shifts ($\lesssim 0.3$ GeV). For $X_t/M_S\sim -2$, $M_h$ is shifted upwards by the inclusion of NNLL resummation by up to $1$ GeV, whereas $M_h$ is shifted downwards by up to $0.5$ GeV for $X_t/M_S =2$. This behavior is mainly caused by the $\mathcal{O}(\alpha_s\alpha_t)$ matching condition of $\lambda$, which exhibits a similar dependence on $\xf$. The large positive shift for negative $X_t$ compensates the downwards shift originating from the electroweak NLO correction to the $\MS$ top-quark mass. This downwards shift is however enhanced by the negative shift for positive $X_t$. This is the reason for the asymmetric behaviour observed in the right panel of Fig.~\ref{FigFHcomp}.

Note that the comparison made in the right panel of Fig.~\ref{FigEWinoNNLLContr} does not exhibit the effect of the two-loop corrections to the $\MS$ top mass, since also for the curve without NNLL resummation the two-loop QCD corrections in the $\MS$-mass -- pole-mass relation are employed. We have kept them because they
constitute the by far dominant part of the two-loop corrections to the $\MS$ top mass, shifting the $\MS$ top mass down by 1.9 GeV. This downwards shift causes a downwards shift in $M_h$ of about the same size, as discussed before in the context of the electroweak NLO corrections to the $\MS$ top mass. Two-loop corrections to the $\MS$ top mass are formally not needed in the case of LL and NLL resummation. This means actually that the main effect of going from NLL to NNLL resummation is caused by the higher-order matching condition of the $\MS$ top mass, as in the case of including electroweak corrections into the resummation procedure.

\section{Conclusions}\label{SecConclusions}

We have presented and discussed the inclusion of electroweak contributions, electroweakino and gluino thresholds, and NNLL resummation in the EFT resummation of logarithmically enhanced terms in the calculation  
of the lightest Higgs boson mass $M_h$, on top of the fixed-order one- and two-loop computation
as currently available in the code \FH.  Special attention is payed to a consistent combination
of fixed-order diagrammatic and EFT methods taking care of scheme conversion and proper subtractions 
to avoid double counting.    
These improvements have become part of {\tt FeynHiggs}.  
They shift the prediction for $M_h$,
especially pronounced for positive values of the stop-mixing parameter $X_t$ 
with downwards shifts in $M_h$ of about $1.7$ GeV.

We found that this is mainly caused by the electroweak NLO corrections to the $\MS$ top-quark mass. The genuine effect of resumming electroweak contributions shifts the Higgs mass upwards compensating the downwards shift induced by the smaller $\MS$ top-quark mass. This effect becomes only relevant for SUSY scales larger than a few TeV. Furthermore, electroweak NLL contributions are found to be much smaller than electroweak LL contributions. 

We also investigated the effect of various intermediate thresholds. In our framework, an electroweakino threshold yields significant contributions only for SUSY scales above 5 TeV. We found that a gluino threshold is completely negligible, since the main contributions sensitive to the gluino mass are already captured by the two-loop Feynman-diagrammatic result.

Furthermore, we found NNLL resummation of $\mathcal{O}(\alpha_s,\alpha_t)$ to shift the lightest Higgs mass downwards for positive stop mixing, whereas it leads to a larger upwards shift for negative values of $X_t$.

We aim to compare the results thoroughly to other publicly available codes \cite{Vega:2015fna,Gabe:2016} in an upcoming publication. We also plan to extend the resummation procedure to scenarios with light non-SM Higgs bosons~\cite{Cheung:2014hya,Lee:2015uza}.

\section*{Acknowledgments}
\sloppy{We are thankful to Thomas Hahn for his invaluable help
  concerning all issues related to \FH, and to  
Sven Heinemeyer and Georg Weiglein for useful discussions.}

\appendix

\section{RGEs for SM with gluinos}\label{GluinoRGE}

The RGEs for the SM with an added gluino are extracted from the RGEs
listed in~\cite{Giudice:2011cg}. The authors of~\cite{Giudice:2011cg}
considered a split model, where all gauginos and higgsinos are assumed to be mass
degenerate. In order to get the gluino part separately we had to disentangle
the gluino and electroweakino contributions in the RGEs of~\cite{Giudice:2011cg}. 
The extracted RGEs have been crosschecked using {\tt SARAH}, version {\tt 4.9} \cite{Staub:2013tta}.

The normalization of $\lambda$ and $v$ is fixed by the following convention for the SM Higgs potential,
\begin{align}
V(\Phi) = -\frac{m^2}{2}\Phi^\dagger\Phi+\frac{\lambda}{2}(\Phi^\dagger\Phi)^2,
\end{align}
with the SM Higgs doublet
\begin{align}
\Phi = \begin{pmatrix} G^+ \\ \frac{1}{\sqrt{2}}(v+h+i G^0)\end{pmatrix}.
\end{align}
Using this convention, the RGEs below and above the gluino threshold are given by
\begin{subequations}
\begin{align}
\frac{dg_3^2}{d\ln Q^2} =& \frac{g_3^4}{(4\pi)^2}\bigg[-\langle 7;5\rangle\bigg] + \frac{g_3^4}{(4\pi)^4} \bigg[-\langle 26;-22\rangle g_3^2 - 2 y_t^2\bigg] ,\\
\frac{dy_t^2}{d\ln Q^2} =& \frac{y_t^2}{(4\pi)^2}\bigg[\frac{9}{2}y_t^2 - 8 g_3^2 \bigg] \nonumber\\
&+ \frac{y_t^2}{(4\pi)^4}\bigg[y_t^2\bigg(-12 y_t^2 - 6 \lambda + 36 g_3^2\bigg) +\frac{3}{2}\lambda^2 - \langle108;\frac{284}{3}\rangle g_3^4\bigg] ,\\
\frac{d\lambda}{d\ln Q^2} =& \frac{6}{(4\pi)^2}\bigg[\lambda^2 + \lambda y_t^2 - y_t^4\bigg] \nonumber\\
&+ \frac{1}{(4\pi)^4}\bigg[ y_t^4\bigg(30 y_t^2 -32 g_3^2\bigg)+\lambda y_t^2 \bigg(40 g_3^2-\frac{3}{2}y_t^2\bigg)-36\lambda^2 y_t^2-39\lambda^3 \bigg] .
\end{align}
\end{subequations}
The notation $\langle a;b\rangle$ indicates that $a$ is to be used for scales below $M_{\tilde g}$ and $b$ for scales above $M_{\tilde g}$. For clarity, we omit terms proportional to the electroweak gauge couplings or the effective Higgs-Higgsino-gaugino couplings, which are not modified by the presence of the gluino.

\section{Explicit one-loop expressions}\label{AppExplicitExpr}

Extracting all one-loop leading logarithms out of the Feynman-diagrammatic result yields
\begin{align}
(M_h^2)^{\text{1L,LL}}=& M_Z^2\, c_{2\beta}^2(M_t)-\frac{1}{72 \pi^2  v^2} \cdot\nonumber\\
& \bigg\{-\frac{3}{8}\bigg[288 m_t^4 + 144 m_t^2 M_Z^2 c_{2\beta} + 296 M_W^2 - 336 M_W^2 M_Z^2 +189 M_Z^4  \nonumber\\
&\hspace{1.1cm} + 4\Big(62 M_W^4 - 84 M_W^2 M_Z^2 + 39 M_Z^4\Big)c_{4\beta} - 9 M_Z^4\, c_{8\beta}\bigg]\ln \frac{M_S^2}{M_t^2} \nonumber \\
&\hspace{.27cm} + 3\bigg[44 M_W^4- 10 M_W^2 M_Z^2 + 11 M_Z^4 + \Big(20 M_W^4 - 10 M_W^2 M_Z^2 - M_Z^4\Big) c_{4\beta} \bigg] \ln \frac{M_\chi^2}{M_t^2} \bigg\},
\end{align}
where $M_Z$ ($M_W$) is the mass of the $Z$ ($W$) boson, $m_t$ is the top-quark mass used to parametrize the diagrammatic result (i.e.\ OS mass or $\MS$ mass) and the abbreviation $c_x \equiv \cos x$ is used. The terms proportional to $ \ln(M_\chi^2/M_t^2)$ originate from charginos and neutralinos. The contributions of all other sectors yield the terms proportional to $\ln(M_S^2/M_t^2)$.

On the other hand, the EFT calculation yields
\begin{align}
(M_h^2)^{\text{1L,LL}}=& M_Z^2\, c_{2\beta}^2(M_S)-\frac{1}{72 \pi^2  v^2} \cdot\nonumber\\
& \bigg\{-\frac{3}{8}\bigg[288 m_t^4 - 144 m_t^2 M_Z^2 c_{2\beta}^2 + 296 M_W^2 - 336 M_W^2 M_Z^2 +189 M_Z^4  \nonumber\\
&\hspace{1.1cm} + 4\Big(62 M_W^4 - 84 M_W^2 M_Z^2 + 39 M_Z^4\Big)c_{4\beta} - 9 M_Z^4 c_{8\beta}\bigg]\ln \frac{M_S^2}{M_t^2} \nonumber \\
&\hspace{.27cm} + 3\bigg[44 M_W^4- 10 M_W^2 M_Z^2 + 11 M_Z^4 + \Big(20 M_W^4 - 10 M_W^2 M_Z^2 - M_Z^4\Big) c_{4\beta} \bigg] \ln \frac{M_\chi^2}{M_t^2} \bigg\}.
\end{align}
Using the dominant one-loop RGE for $\tan\!\beta$ as explained in Section~\ref{SecCombining},
\begin{align}
c_{2\beta}^2(M_S) = c_{2\beta}^2(M_t) + \frac{3}{2\pi^2}\frac{m_t^2}{v^2} c_\beta^2\, c_{2\beta} \ln\frac{M_S^2}{M_t^2} + ...,
\end{align}
we recover the result of the diagrammatic calculation.

\newpage

\bibliographystyle{JHEP}
\bibliography{bibliography}{}

\end{document}